\documentclass[11pt,a4paper]{article}
\pdfoutput=1
\usepackage{jheppub}
 \newcommand{\PRE}[1]{}
\usepackage{amsmath,amsfonts,epsfig,graphicx}
\usepackage{multirow}
\usepackage{indentfirst}
\usepackage[latin1]{inputenc}
\usepackage{amssymb}
\usepackage{verbatim}
\usepackage{float}
\usepackage{subfig}
\usepackage{slashed} 
\usepackage{siunitx}
\usepackage{hyperref}
\usepackage{todonotes}
\usepackage{lineno}

\newcommand{\pythia}{{\sc Pythia}}

\newcommand{\delphes}{{\sc Delphes}}

\newcommand{\al}{{\it et al. }}
\newcommand{\MG}{M{\scriptsize AD}G{\scriptsize RAPH}5\_{\scriptsize aMC@NLO}}

\begin{document}

\title{A Comparative Study of Z$^{\prime}$ mediated Charged Lepton Flavor Violation at future lepton colliders}

\author[a]{Jingshu Li,}
\author[a]{Wanyue Wang,}
\author[a]{Xunye Cai,}
\author[a]{Chuxue Yang,}
\author[a]{Meng Lu,}
\author[a]{Zhengyun You,}
\author[b]{Sitian Qian,}
\author[b]{and Qiang Li}

\affiliation[a]{
School of Physics, Sun Yat-Sen University, Guangzhou 510275, China}
\affiliation[b]{
School of Physics and State Key Laboratory of Nuclear Physics and Technology, Peking University, Beijing, 100871, China}
\emailAdd{lijsh53@mail2.sysu.edu.cn}
\emailAdd{wanggy25@mail2.sysu.edu.cn}
\emailAdd{caixy67@mail2.sysu.edu.cn}
\emailAdd{yangchx2667@mail2.sysu.edu.cn}
\emailAdd{lumeng5@mail.sysu.edu.cn}
\emailAdd{youzhy5@mail.sysu.edu.cn}
\emailAdd{sitian.qian@cern.ch}
\emailAdd{qliphy0@pku.edu.cn}

\abstract{Charged lepton flavor violation (CLFV) represents a transition between charged leptons of different generations that violates lepton flavor conservation, which is a clear signature of possible new physics beyond the standard model. By exploiting a typical example model of extra Z$^{\prime}$ gauge boson, we perform a detailed comparative study on CLFV searches at several future lepton colliders, including a 240 GeV electron-positron collider and a TeV scale muon collider. Based on detailed signal and background Monte-Carlo studies with fast detector simulations, we derive the potentials in searching for Z$^{\prime}$ mediated CLFV couplings with $e\mu$, $e\tau$ and $\mu\tau$ of different future colliders. The results are compared with the current and prospect limits set by either low-energy experiments or the high-energy LHC experiments. The sensitivity of the $\tau$ related CLFV coupling strength at future lepton colliders will be significantly improved in comparison to the current best constraints and the prospect constraints for the $\mu\tau$ channel.
}

\maketitle

\section{Introduction}
The standard model (SM) has been proved to be a theory with great success during the past decades, especially after the discovery of Higgs boson, the last piece of the puzzle. The lepton flavor is supposed to be conserved during interactions within the SM, which  allows the charged lepton flavor violation (CLFV) neither at tree-level nor at one-loop level. However, the discovery of neutrino mass and neutrino oscillations enable the loop level CLFV within the SM, although it is highly suppressed by a factor of $(\Delta m_{ij}/M_{W})^4$ due to the tiny mass of neutrinos, e.g., the branching ratio of $\mu \rightarrow e\gamma$ decay reads:
\begin{equation}
    {\cal B}(\mu \to e \gamma )=\frac{3\alpha}{32 \pi}\left|\sum_{i=2,3} 
    U_{\mu i}^*U_{ei} \frac{\Delta m_{i1}^2}{M_W^2}\right|^2 \sim 10^{-54}\,, 
\end{equation}
where $\alpha$ denotes the fine structure constant, $U_{\mu i}$ is the element of the neutrino mixing matrix, $\Delta m_{i1}^2$ is the difference of the squared neutrino masses, and $M_W$ is the mass of the W boson. This branching ratio is obviously a value far away from what can be measured experimentally, which means any observed signature of $\mu \rightarrow e\gamma$ is definitely a discovery of new physics beyond the SM (BSM). In addition to the $\mu \rightarrow e\gamma$ channel, muon related CLFV is also performed in $\mu^- + N \rightarrow e^- + N$ and $\mu \rightarrow 3e$ channels, the history of the CLFV searches are summarized in Ref.~\cite{Bernstein:2013hba, Davidson:2022jai,Calibbi:2017uvl}. 

The search for CLFV has attracted long standing interest since it has great potential to probe new physics indirectly at energy scales much higher than what is going to be accessible by the colliders in the foreseeable future. Various searches have been performed at several different approaches~\cite{Workman:2022ynf}, including the muon-based experiments (such as $\mu \to e \gamma$ at MEG-II~\cite{Renga:2022tex,MEGII:2021fah}, $\mu \to e e e$ at Mu3e~\cite{Mu3e:2020gyw,Dittmeier:2022nwi} and $\mu^{-}N \to e^{-}N$ at Mu2e~\cite{Mu2e:2014fns} and COMET~\cite{COMET:2018auw}), B factories, and high energy colliders such as LEP and LHC, targeting at the CLFV decays of meson\cite{cleo3:2008,BaBar:2021loj,BESIII:2022exh,SND:2010}, $\mu$, $\tau$, Z~\cite{ATLAS:2014vur,OPAL:1995grn,DELPHI:1996iox} and Higgs boson~\cite{ATLAS:2019pmk,CMS:2017con,ATLAS:2019old}.


In the next decades, the LHC and the High-Luminosity LHC (HL-LHC), together with other future colliders in design, will further explore the SM and search for BSM physics. The majority of the proposed future machines are lepton colliders, designed primarily for Higgs-boson precision measurements. The most promising proposals include linear or circular electron-positron colliders~\cite{ILC, FCC, CEPC, CLIC} and muon colliders~\cite{Muc}. In this study, we are interested in searching for CLFV at those future lepton colliders. Similar studies~\cite{Li:2018cod,Li:2021lnz,Homiller:2022iax} exist in literature but either only for electron positron colliders or being performed with models (e.g., effective operators) different from this study.

Many BSM models enhancing CLFV effects up to a detectable level have been proposed, including supersymmetry~\cite{SUSY}, heavy Z$^{\prime}$~\cite{Leike, HeavyZ, Altmannshofer:2022fvz, Buras:2021btx} and scalar neutrinos in R-parity-violating~\cite{Rparity}. A new $U(1)$ gauge symmetry resulting in a massive neutral vector boson known as a  Z$^{\prime}$ boson is one of the common extensions of the SM~\cite{Leike, HeavyZ,Erler:2009jh, Carena:2004xs}, and Z$^{\prime}$ can have interaction with a pair of W bosons~\cite{2015jvn}.

The search presented in this paper assumes a Z$^{\prime}$ boson with the same quark couplings and chiral structure as the SM Z boson~\cite{Fox:2011qd, Accomando:2010fz, Komachenko:1989qn, Crivellin:2015era}, but with different mass and allowing CLFV couplings, similarly as done previously in an ATLAS study~\cite{ATLAS:2018mrn}. Such Z$^{\prime}$ interacts with charged leptons, of which the coupling strengths can be described phenomenologically by the coupling matrix as shown in Equation~\ref{eq:coupling_full}, in which $\lambda_{ij}$ is the CLFV coupling of lepton $i$ and lepton $j$. Only one CLFV coupling $\lambda_{ij} (i \neq j)$ is assumed to be non-zero (set as 1) at any time for the purpose of setting the upper limits, while the diagonal couplings $\lambda_{ll} (l=e, \mu, \tau)$ are always set as 1.

\begin{equation}
{\lambda_{ij}} =  
\begin{pmatrix} 
\lambda_{ee}&\lambda_{e\mu}&\lambda_{e\tau}\\
\lambda_{\mu e}&\lambda_{\mu\mu}&\lambda_{\mu\tau}\\
\lambda_{\tau e}&\lambda_{\tau \mu}&\lambda_{\tau\tau} 
\end{pmatrix}.
\vspace{0.5cm}
\label{eq:coupling_full}
\end{equation}

Below we also compare our results with the theoretical constraints from low energy muon based experiments. Take the $\mu - e$ conversion as an example, the CLFV coupling $\lambda_{e\mu}$ can be transformed from the branching ratio $R$ for $\mu - e$ conversion in nuclei~\cite{1993ta} 

\begin{align}
\lambda^{2}_{e\mu}=&\frac{2\pi^{2}\Gamma_{\rm capture}ZR}{G^{2}_{F}\alpha^{3}m^{5}_{\mu}Z^{4}_{eff}\lvert F(q)\rvert ^{2}}\frac{M^{4}_{Z^{\prime}}}{M^{4}_{Z}}\times\frac{1}{s^{4}_{W}+(s^{2}_{W}-\frac{1}{2})^{2}} \nonumber \\
&\times\frac{1}{[(2Z+N)(\frac{1}{2}-\frac{4}{3}s^{2}_{W})+(Z+2N)(-\frac{1}{2}+\frac{2}{3}s^{2}_{W})]^{2}}.
\label{eq:lamemu}
\end{align}
The branching ratio $R$ is normalized to the total nuclear muon capture rate $\Gamma_{\rm capture}$ measured experimentally with good precision, the limit $R$ is taken from PDG by SINDRUMII experiment corresponding to the experiment using $^{197}_{79}$Au or $^{48}_{22}$Ti as the target~\cite{SINDRUMII:2006dvw}. $\Gamma_{\rm capture}$ is calculated by the muon captured lifetime $\tau$ on target, $G_{F}$ is the Fermi constant, $\alpha$ is the fine structure constant, $m_{\mu}$ is the rest mass of muon, $Z_{eff}$ and F(q) are nuclear parameters~\cite{Suzuki:1987jf}, $Z$ is the atomic number, $N$ is the number of neutrons in the nucleus, and $s_{W}$ is the sine of the weak mixing angle, the values are listed in Table~\ref{tab:eqlam}. 
Using Equation~\ref{eq:lamemu}, the constraints on the CLFV coupling from $\mu-e$ conversion can be obtained. Other low energy limits, such as those from $\mu \to e\gamma$ and $\mu \to eee$, are taken from Ref.~\cite{zpcoupl}. For comparison with the next generation $\mu - e$ conversion experiments, the prospect upper limit with $^{27}_{13}$Al target in Mu2e~\cite{Mu2e:2014fns} or COMET~\cite{COMET:2018auw} experiment is also listed.
For the $\tau$ channel, assuming that $\tau \to e\gamma$ is similar to $\mu \to e\gamma$, we use the same method to get the limits, except for replacing $\mu$ with $\tau$ and using the approximation that $m_{\tau} \gg m_{e}$.

\begin{table}[!hbtp] 
\setlength{\abovecaptionskip}{0.05cm}
	 \caption{The parameters for different targets in $\mu - e$ conversion Equation~\ref{eq:lamemu}.}
\begin{center}
   \begin{tabular}{ccccccc}
	 \hline\hline
   Target &  R &   $\tau$ & $Z_{eff}$ & F(q) & Z & N 
   \\
	    \hline
    Ti  & $<4.3\times 10^{-12}$~\cite{SINDRUMII:1993gxf} &  329.3 ns & 17.60 & 0.54 &22 & 26
    \\
    Au  & $<7\times 10^{-13}$~\cite{SINDRUMII:2006dvw} & 72.6 ns & 33.64 & 0.13 &79 & 118
    \\
    Al  & $<2.9\times 10^{-17}$(prospect)~\cite{Mu2e:2014fns} & 864 ns & 11.48 & 0.67 &13 & 14

     \\
	 \hline\hline
	 \end{tabular}
	 \setlength{\belowcaptionskip}{-0.05cm} 
	 \label{tab:eqlam}
	 \end{center}
	  \end{table}

\section{Sample and analysis}

In this paper, we focus on CLFV searches at a 240 GeV circular electron positron collider (CEPC), and a 6 or 14 TeV muon collider.

\subsection{Signal and background processes}

\begin{figure}[htbp!]
    \centering
    \subfloat[\label{fig:emu}]{\includegraphics[width=0.32\textwidth]{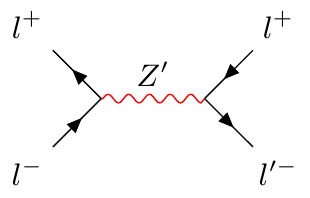}}\qquad
    \subfloat[\label{fig:etau}]{\includegraphics[width=0.32\textwidth]{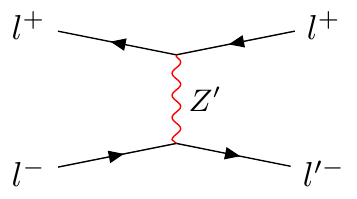}}\qquad

    \caption{Example Feynman diagrams at CEPC and muon colliders for CLFV  s-channel (a) and t-channel (b) processes, mediated by Z$^\prime$.}
    \label{fig:Feynman}
\end{figure}

Possible signal processes include $e e\to e \mu$, $e e \to e \tau$, $\mu \mu \to e \mu$ and $\mu \mu \to \mu \tau$. Figure~\ref{fig:Feynman} shows example Feynman diagrams for CLFV processes mediated through a Z$^\prime$ boson as mentioned above. The main background processes are summarized in Table~\ref{tab:bkg}. Simulated events are generated corresponding to CEPC with a collision energy at 240 GeV and an integrated luminosity of 5 ab$^{-1}$, or a muon collider with collision energies of 14 (6) TeV and an integrated luminosity of 4 ab$^{-1}$. 

\begin{table}[!hbtp] 
\setlength{\abovecaptionskip}{0.05cm}
	 \caption{Summary of the main background processes.}
\begin{center}
   \begin{tabular}{cc}
	 \hline \hline
   Signal process        &  Selected background
   \\
	    \hline
     $e e\to e \mu$ & WW,\,H$\nu\bar{\nu}$(H$\to \tau\tau$),\,H$\nu\bar{\nu}$(H$\to$ WW),\,$\tau\tau$
     \\
          $e e\to e \tau$ & WW,\,WW$\nu\bar{\nu}$,\, H$\nu\bar{\nu}$(H$\to \tau\tau$),\,$\tau\tau$
     \\
          $\mu \mu \to e \mu$ & WW,\,WW$\nu\bar{\nu}$,\, H$\nu\bar{\nu}$(H$\to \tau\tau$),\,H$\nu\bar{\nu}$(H$\to$WW),\,$\tau\tau$
     \\
          $\mu \mu \to \mu \tau$ & WW,\,WW$\nu\bar{\nu}$,\, H$\nu\bar{\nu}$(H$\to \tau\tau$),\,H$\nu\bar{\nu}$(H$\to$WW),\,$\tau\tau$
     \\
	 \hline\hline
	 \end{tabular}
	 \setlength{\belowcaptionskip}{-0.05cm} 
	 \label{tab:bkg}
	 \end{center}
	  \end{table}

\subsection{Event generation and simulation}
Both signal and background events are generated with \MG~\cite{MG5} (MG5aMC) version 3.1.1, then showered and hadronized by \pythia8~\cite{pythia8}. Specially for the CEPC, the initial-state radiation (ISR) effect~\cite{ISR} is included as well. The detector effects are simulated using~\delphes~\cite{delphes} version 3.5 with the default cards for the corresponding collider detectors.  

We describe the selection criteria as below. First, the events must include exactly two leptons with transverse momentum $p_T > 10$ GeV and absolute pseudo-rapidity $|\eta| < 2.5$, and satisfy the requirements of lepton flavor and charge conversation from the Z$^\prime$ boson decay, i.e., for $e^{+} e^{-} \to e^{+} \mu^{-} (e^{-} \mu^{+})$, all signal and background events are required to have only one $e^{+} (e^{-})$ and one $\mu^{-} (\mu^{+})$. For the final state containing $\tau$ which is reconstructed within the jet collection in \delphes, the requirements of the jets are $p_T > 20$ GeV and $|\eta| < 5$. 

The $\mu$ tracking efficiency is assumed to be 100$\%$ for CEPC with $0.1 < |\eta| \leq 3$, and 0$\%$ for the rest area. At the muon collider, the $\mu$ tracking efficiency is listed in Table~\ref{tab:eff}.  For the $\tau$ channel, the $\tau$ tagging efficiency is 40$\%$ for the CEPC and 80$\%$ for the muon collider with $p_T > 10$ GeV, as defined in \delphes~cards~\cite{delphes}.
 \begin{table}[!hbtp] 
\setlength{\abovecaptionskip}{0.05cm}
	 \caption{Summary of $\mu$ tracking efficiency at the muon collider.}
\begin{center}
   \begin{tabular}{cc}
	 \hline \hline
        Conditions &  Efficiency
   \\
	    \hline
     $|\eta| \leq 2.0$ and 0.5 GeV $< p_T \leq 1$ GeV & 95$\%$
     \\
          $|\eta| \leq 2.0$ and $p_T > 1$ GeV & 99$\%$
     \\
          $2.0 < |\eta| < 2.5$ and 0.5 GeV $< p_T \leq 1$ GeV & 90$\%$
     \\
          $2.0 < |\eta| < 2.5$ and $p_T > 1$ GeV & 95$\%$
     \\
               $|\eta| > 2.5$ & 0$\%$
     \\
	 \hline\hline
	 \end{tabular}
	 \setlength{\belowcaptionskip}{-0.05cm} 
	 \label{tab:eff}
	 \end{center}
	  \end{table}

\begin{figure}[htbp!]
    \centering
        \subfloat[\label{fig:eeemumll}]{\includegraphics[width=0.47\textwidth]{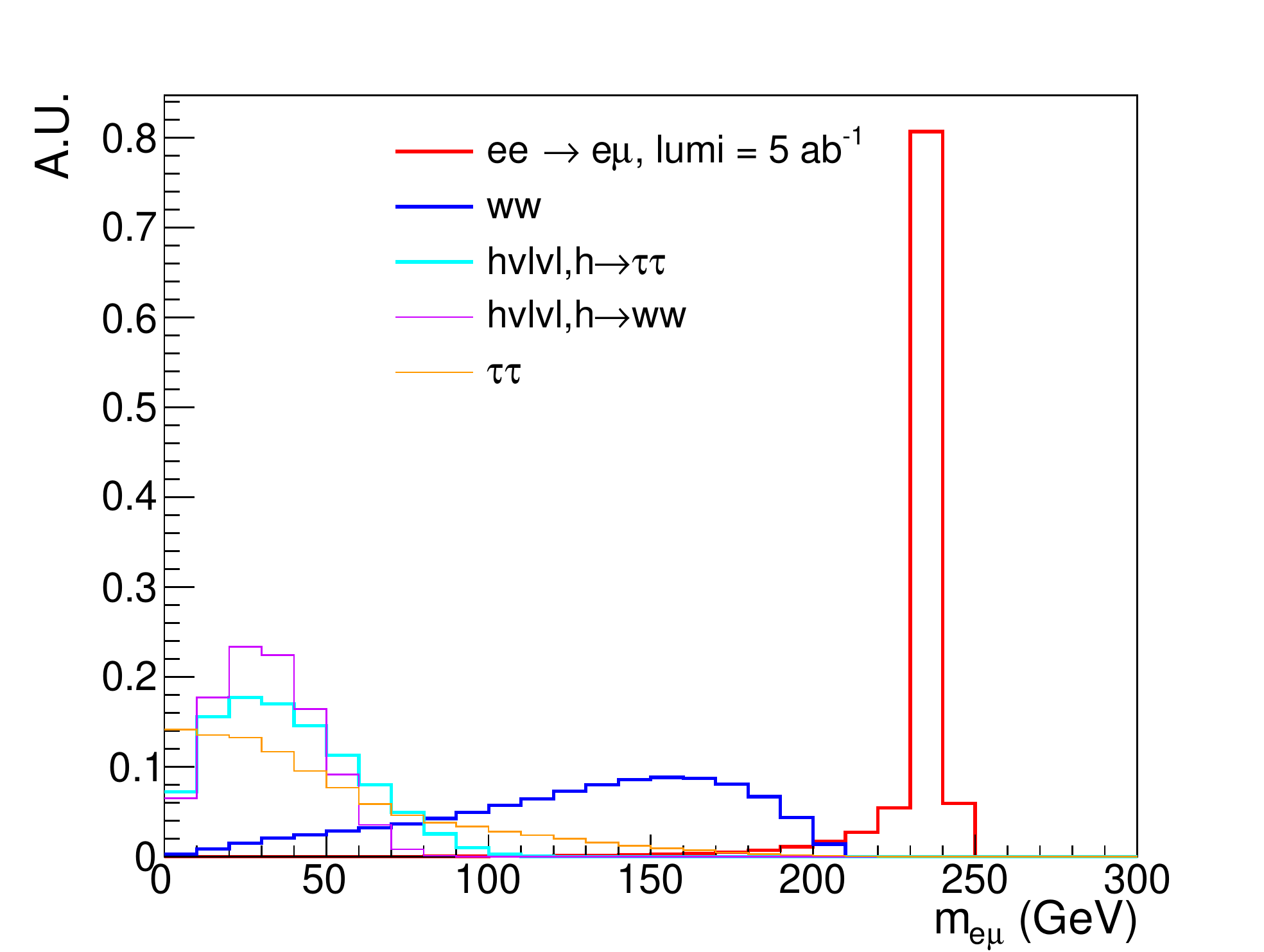}}\qquad
    \subfloat[\label{fig:6muemll}]{\includegraphics[width=0.47\textwidth]{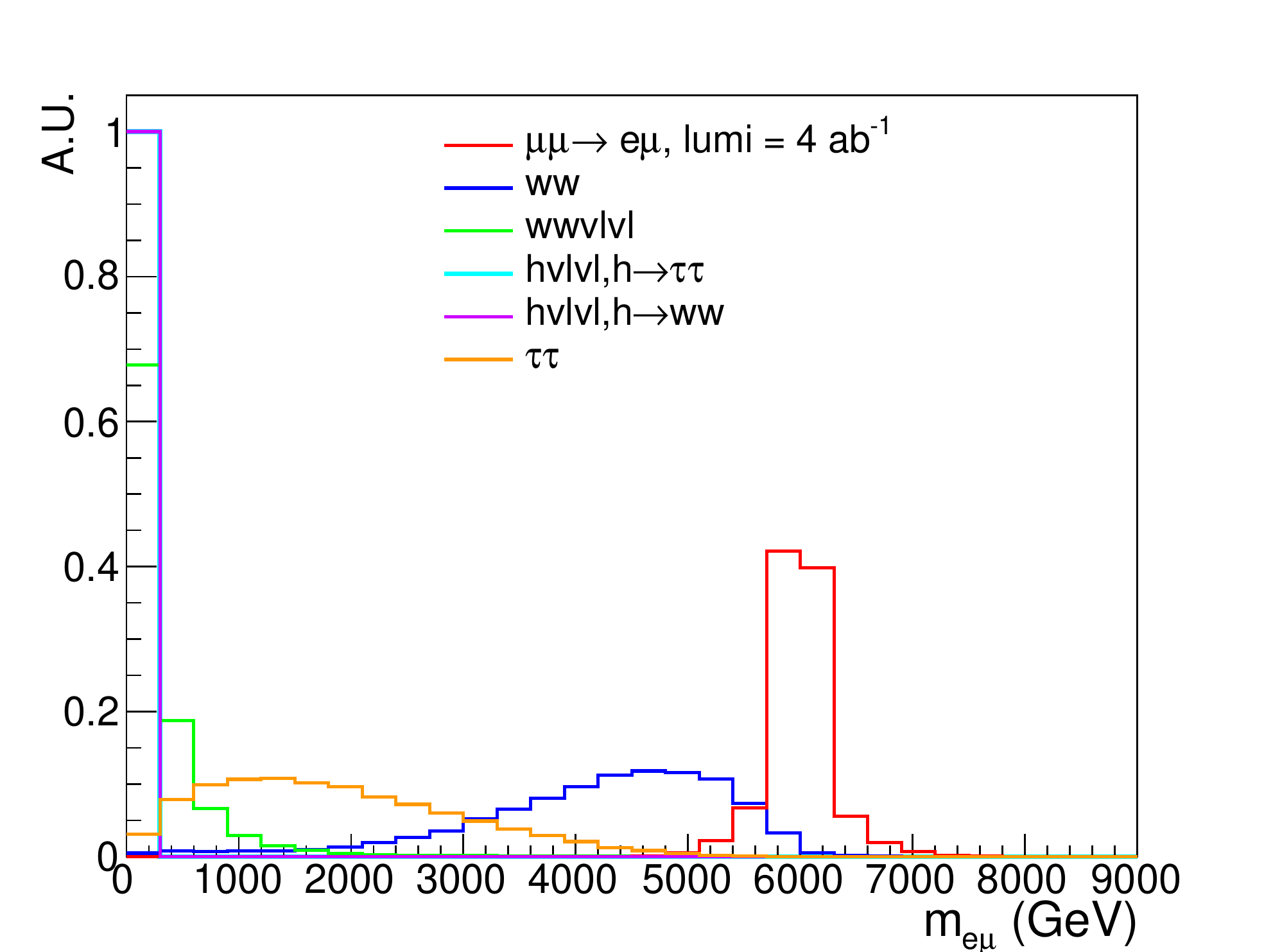}}\qquad
    \subfloat[\label{fig:etaumll}]{\includegraphics[width=0.47\textwidth]{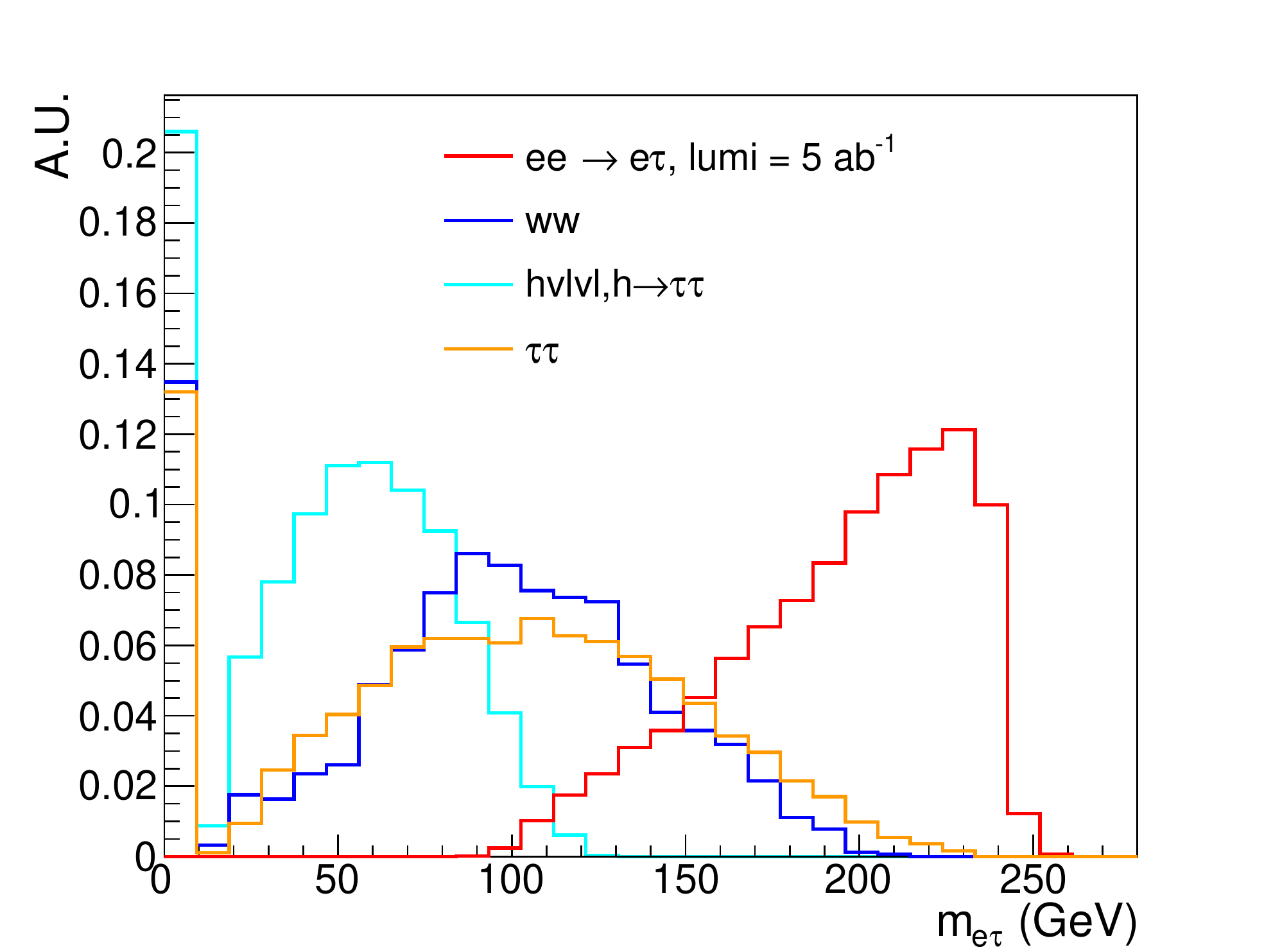}}\qquad
    \subfloat[\label{fig:6mutaull}]{\includegraphics[width=0.47\textwidth]{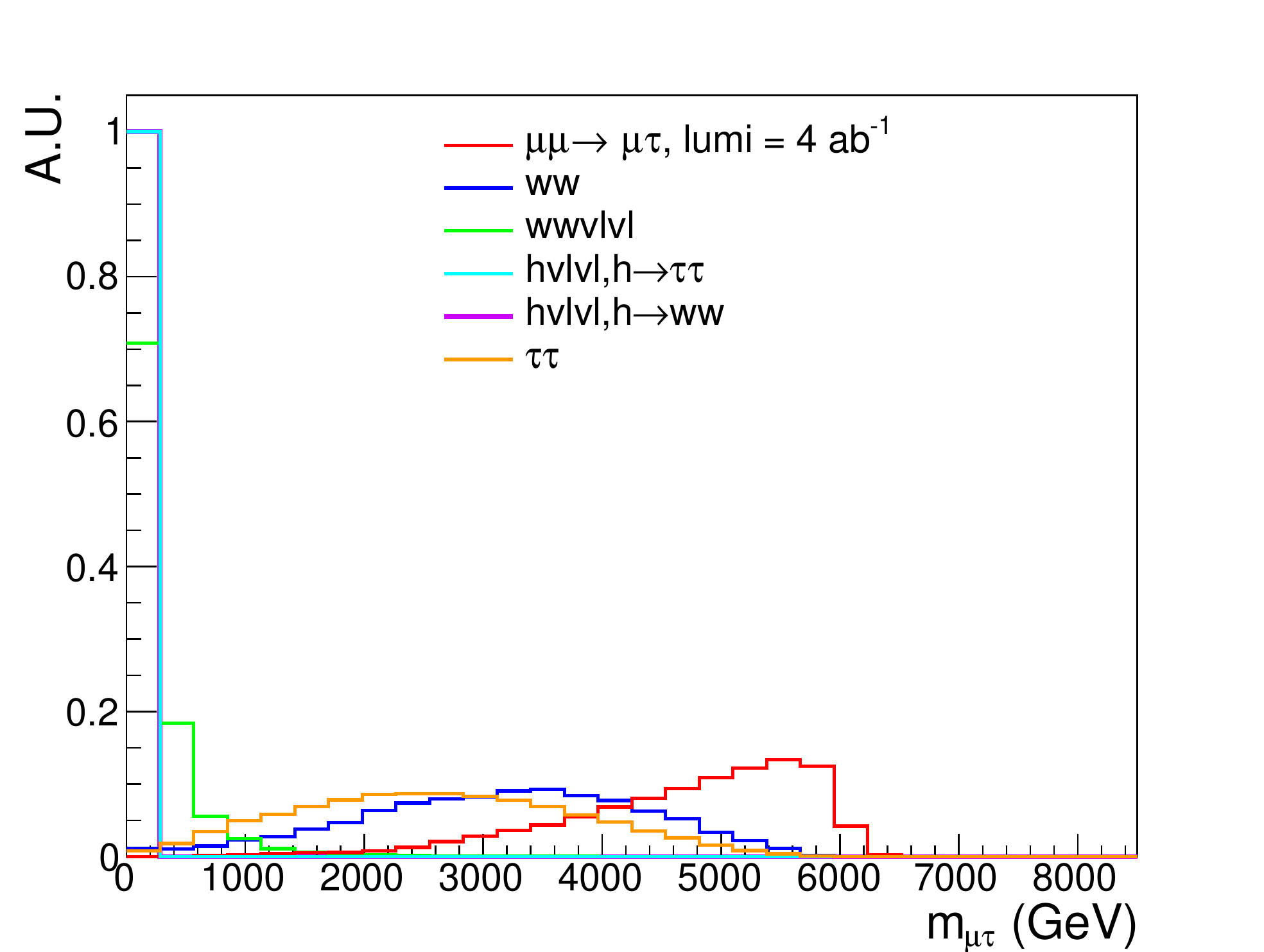}}\qquad

    \caption{Invariant mass distributions of the $e\mu$ channel at CEPC (a) and a 6 TeV muon collider (b), $e \tau$ at CEPC (c) and $\mu \tau$ at a 6 TeV muon collider (d), where the solid red histograms are signals representing Z$^{\prime}$ CLFV mediated processes with the couplings of $e\mu$, $e \tau$ and $\mu \tau$.}
    \label{fig:mll}
\end{figure}

Furthermore, we exploit various physical quantities to separate the signal from the backgrounds. Figure~\ref{fig:mll} shows the invariant mass distributions of final state di-leptons of signal, in comparison with all backgrounds. Compared with the $e\mu$ channel, the invarint masses distribution shape of $\tau$ signal channel is relatively widely peaked around the Z$^\prime$ signal mass, which is due to lower efficiency and larger energy smear of $\tau$ lepton reconstruction. The invariant masses cut is selected at the maximum $S/\sqrt{S + B}$ as the filtering condition, where $S$ denotes the signal events and $B$ denotes the background events. The invariant masses of $e \mu$ (Figure~\ref{fig:eeemumll}) and $e \tau$ (Figure~\ref{fig:etaumll}) at CEPC, $e \mu$ (Figure~\ref{fig:6muemll}) and $\mu \tau$ (Figure~\ref{fig:6mutaull}) at 6 TeV muon collider, as well as the cases at 14 TeV muon collider, are then optimized and required to be greater than 220 GeV, 160 GeV, 5.2 TeV, 4 TeV, 10 TeV and 9.5 TeV, respectively, to maximize signal sensitivities from the backgrounds.

\subsection{Analysis framework}

 After applying all the event selections as mentioned above, we get binned histograms on the final state lepton $p_T$ distributions, which will be exploited to set the upper limits on CLFV couplings. 
 
 The per-event weight is applied to account for the cross-section difference between the processes, the weight is defined by $n_{L_{X}}= \sigma_{X}L/N_{X}$, where $\sigma_{X}$ denotes the cross-section of a process $X$, $L$ denotes the default target integrated luminosity in this study, which is 5 ab$^{-1}$ for CEPC and 4 ab$^{-1}$ for a muon collider, and $N_{X}$ denotes the number of generated events. The signal yields and backgrounds are reweighted to get matched.
 
 
 The test statistics $Z$ is defined as in Equation~\ref{eq:SF_Stats}, for both 95\% CL exclusion limit and 5$\sigma$ discovery limit, where $i$ refers to the bin number, $b$ is the SM background, $n:=s+b$ is the total yields containing both signal and background, $s$ is the beyond SM signal. Both $Z$ statistics subject to $\chi^2$ distribution with the number of degrees of freedom corresponding to the number of bins~\cite{asymptotic}. 

\begin{equation}
\begin{aligned}
    &Z = \sum_{i=1}^{bins} Z_i, \\
    &\begin{cases}
    Z_i := 2\left[ n_i - b_i + b_i\ln(b_i/n_i)\right]&\text{95\% C.L. Exclusion}\\
    Z_i := 2\left[ b_i - n_i + n_i\ln(n_i/b_i)\right]&\text{$5\sigma$ Discovery}.
    \end{cases}
\end{aligned}
    \label{eq:SF_Stats}
\end{equation}

The corresponding signal and background yields are calculated, and the corresponding $Z$ statistics are constructed for each case following Equation~\ref{eq:SF_Stats}. The $Z$ statistic subjects to a $\chi^2$ distribution with 1 degree of freedom.

\section{Results}
\subsection{CLFV study at CEPC}

\begin{figure}[htbp!]
    \centering
    \subfloat[\label{fig:emu1}]{\includegraphics[width=0.472\textwidth]{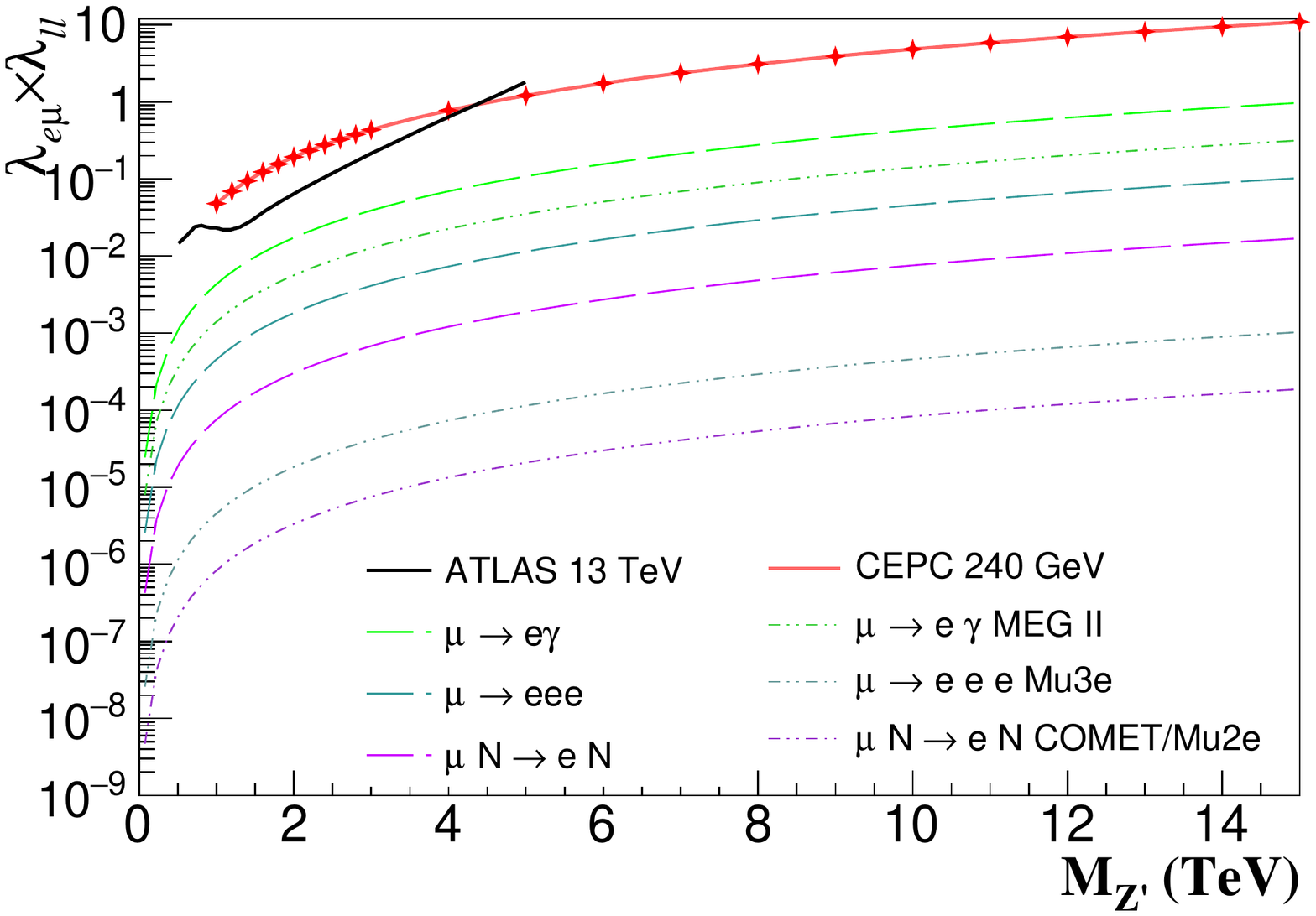}}\qquad
    \subfloat[\label{fig:etau1}]{\includegraphics[width=0.472\textwidth]{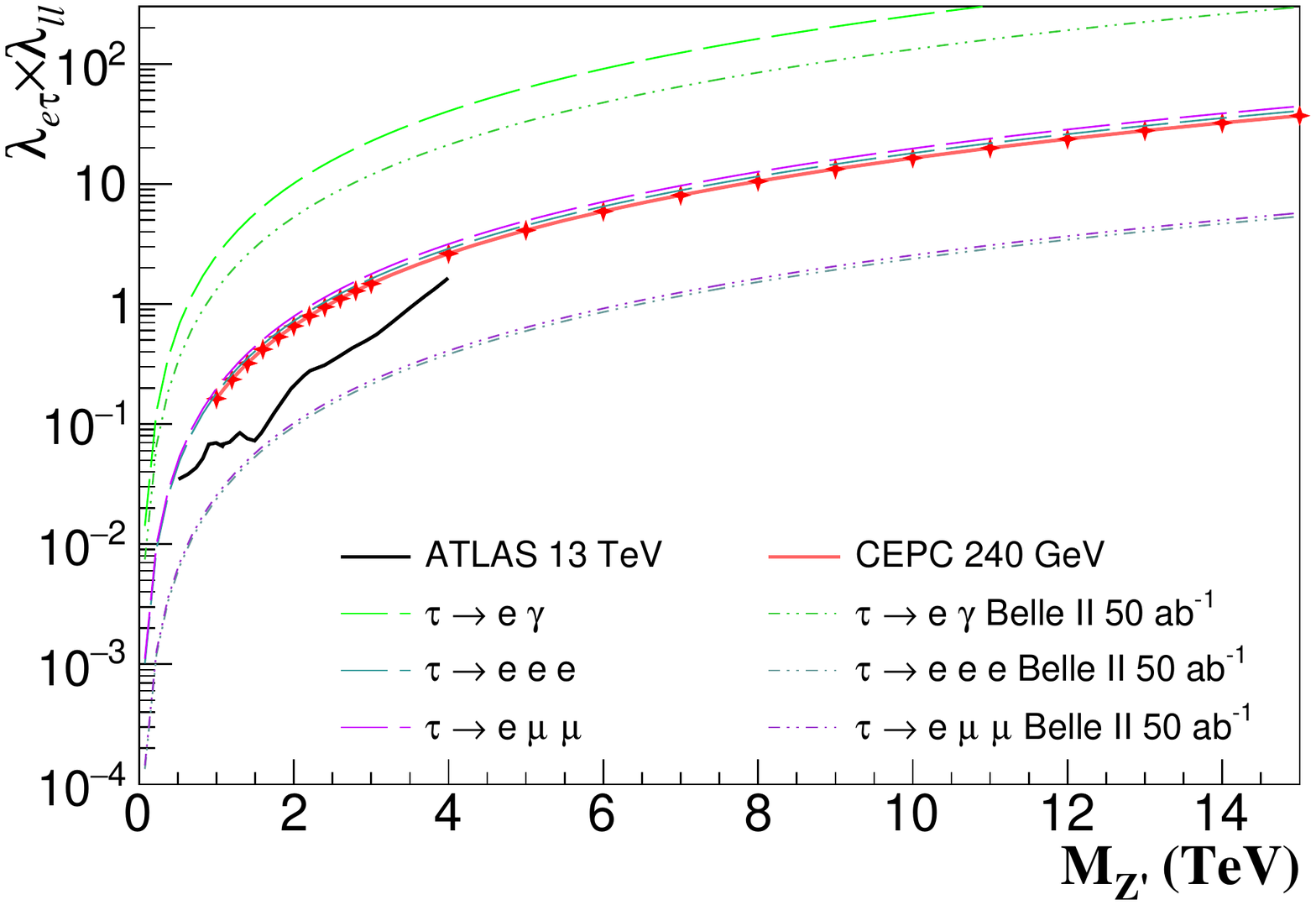}}\qquad

    \caption{95\% C.L. exclusion lines of CLFV on the couplings $\lambda_{e \mu}$ (a) and $\lambda_{e \tau}$ (b) products the diagonal coupling $\lambda_{ll}$ at CEPC ((red line) and ATLAS experiment (black line). The curves are plotted as functions of M$_{Z^{\prime}}$ from the cross-section times branching ratio limits. The exclusion lines with the current low-energy experiments (dashed lines) and future experiments (dash-dotted lines) are also plotted.}
    \label{fig:CEPCret}
\end{figure}

The 95$\%$ confidence level (C.L.) exclusion lines at CEPC and ATLAS, as well as the current and  prospect experimental limits from low-energy $\mu$ and $\tau$ experiments, are converted to the coupling limits $\lambda_{e \mu}\times\lambda_{ll}$ and $\lambda_{e \tau}\times\lambda_{ll}$ using the formula in Ref.~\cite{zpcoupl} respectively for comparison. The results are shown in Figure~\ref{fig:CEPCret}, where the curve trend obtained from the current and prospect experimental limits are similar to the results of the low-energy experiments. 

The most stringent current coupling limits for the $e\mu$ channel are from $\mu$-$e$ conversion and $\mu \to e e e$ in low-energy experiments, and from the ATLAS 13 TeV collision result for the $e\tau$ channel. With the future experiments included, the most stringent coupling will be $\mu$-$e$ conversion from COMET and Mu2e for the $e\mu$ channel, and $\tau$ decays from Belle II~\cite{Banerjee:2022vdd} for the $e\tau$ channel.

\subsection{CLFV study at a Muon Collider}

\begin{figure}[htbp!]
    \centering
    \subfloat[\label{fig:emu2}]{\includegraphics[width=0.472\textwidth]{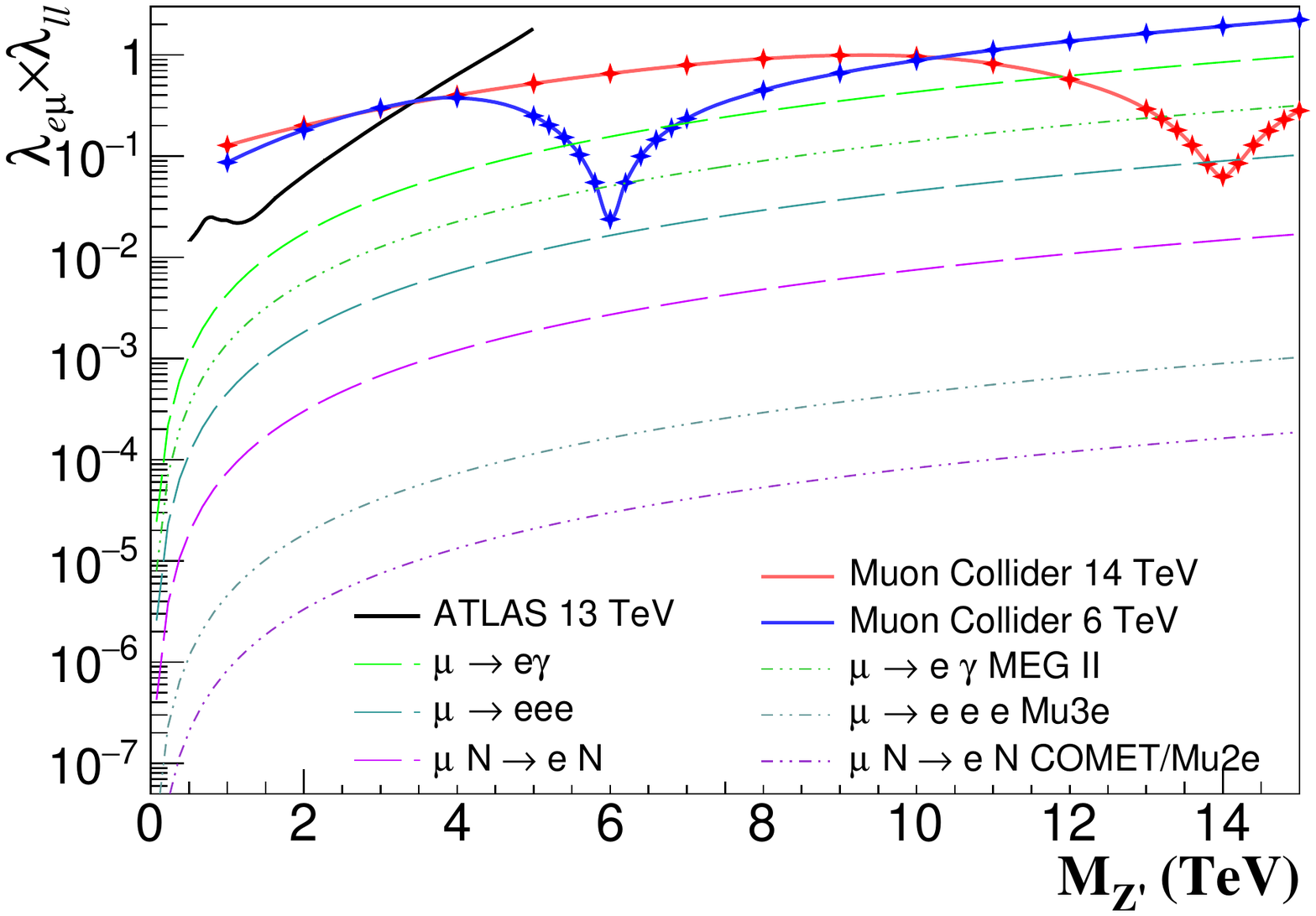}}\qquad
    \subfloat[\label{fig:etau2}]{\includegraphics[width=0.472\textwidth]{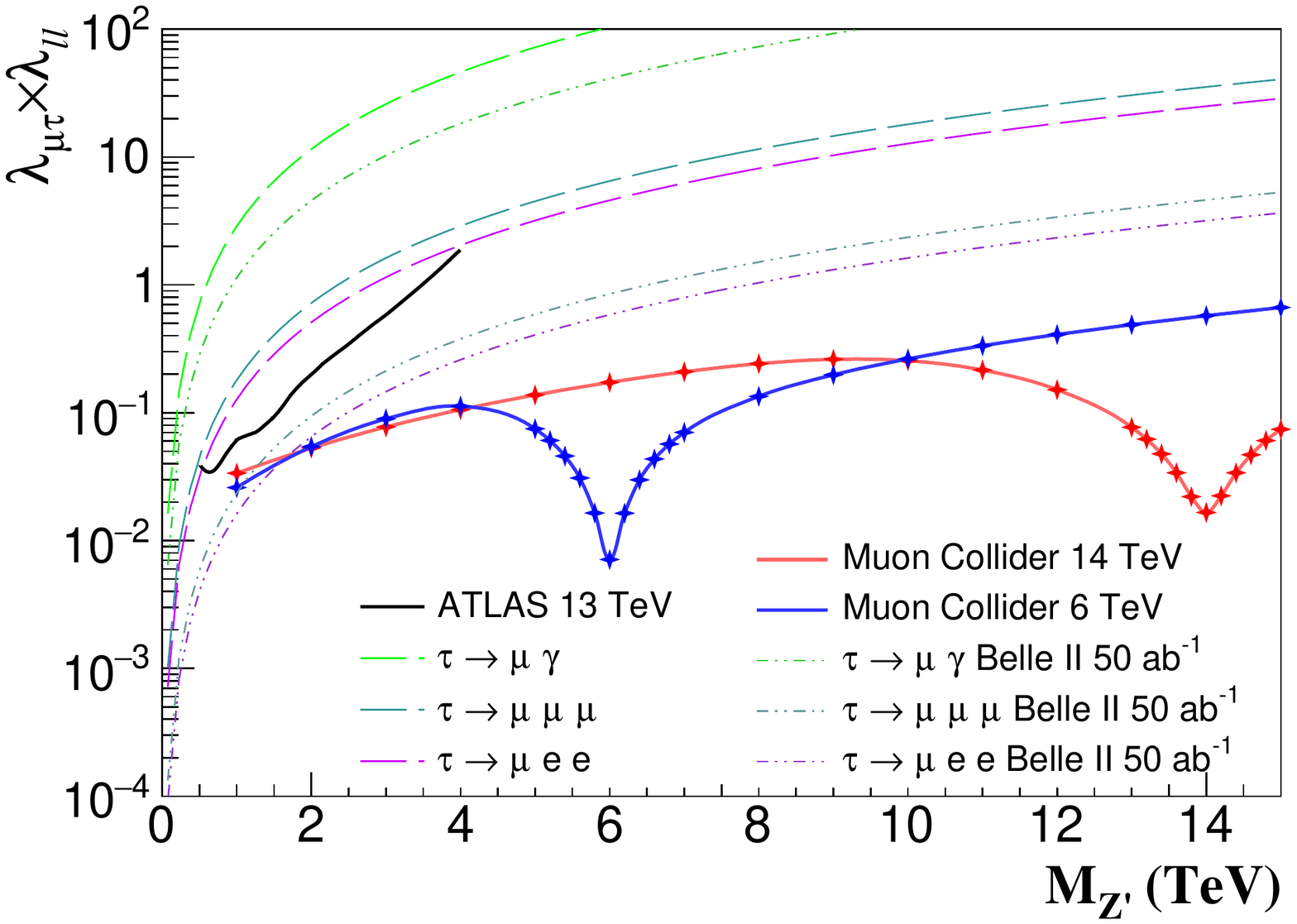}}\qquad

    \caption{95\% C.L. exclusion lines on the couplings $\lambda_{e \mu}$ (a) and $\lambda_{\mu \tau}$ (b) products the diagonal coupling $\lambda_{ll}$ at a 6 TeV (red line) and 14 TeV (blue line) muon collider, as well as ATLAS experiment (black line). The curves are plotted as functions of M$_{Z^{\prime}}$ from the cross-section times branching ratio limits. The exclusion lines with the current low-energy experiments (dashed lines) and future experiments (dash-dotted lines) are also plotted.}
    \label{fig:6mucret}
\end{figure}

Muon collider could be more powerful than CPEC with cleaner environment and higher center of mass. The 95$\%$ C.L. exclusion lines from the $\mu \mu \to e \mu$ and $\mu \mu \to \mu \tau$ processes at 6 and 14 TeV muon collider are converted to coupling limits $\lambda_{e \mu}\times\lambda_{ll}$ and $\lambda_{\mu \tau}\times\lambda_{ll}$ using the formula in Ref.~\cite{zpcoupl} respectively, as shown in Figure~\ref{fig:6mucret}. The constraints from the current and prospect experimental limits from the low-energy $\mu$ and $\tau$ experiments are also included for comparison. 

The current most stringent coupling limits for the $e\mu$ channel are from $\mu$-$e$ conversion in the low-energy experiments, but our results also have stringent constraints at larger Z$^{\prime}$ masses. When the mass of Z$^{\prime}$ is 14 TeV on the 14 TeV Muon collider, the constraints are more stringent than other existing results except for $\mu$-$e$ conversion. 

In comparison to the current results,  the two coupling limits of the $\mu \tau$ channel in this work are the most stringent, and when Z$^{\prime}$ mass is greater than 10 TeV, the upper limit on the 14 TeV Muon collider is more stringent. Including the prospects, the most stringent coupling will be $\mu$-$e$ conversion from COMET and Mu2e for the $e\mu$ channel. While for the $\mu \tau$ channel, the two coupling limits in this work are still the most stringent  when the mass of Z$^{\prime}$ is greater than 1.5 TeV.

The strongest constraint on $\mu \tau$ coupling is from the 6 TeV Muon collider, which reaches the magnitude of 10$^{-3}$ when the mass of Z$^{\prime}$ is 6 TeV, and is stronger than all the existing and most of the prospect CLFV limits. 

\section{Outlook and conclusion} 

 By exploiting a typical example model of extra Z$^{\prime}$ gauge boson, we perform a detailed comparative study on CLFV searches at several future colliders, including a 6 (14) TeV scale muon collider and a 240 GeV electron-positron collider. Based on the event generator software MadGraph, PHTHIA and fast detector simulation implemented with the Delphes framework, we compare the potentials for future lepton colliders to probe Z$^{\prime}$ CLFV couplings with either $e\mu$, $e\tau$ or $\mu\tau$, through the processes of $e e \to e \mu$, $e e \to e \tau$, $\mu \mu \to e \mu$ and $\mu \mu \to \mu \tau$. The upper limits at the 95$\%$ C.L. are set at different Z$^{\prime}$ masses. We find that the sensitivity on the $\tau$ related CLFV coupling strength can be significantly improved in comparison to the current best constraints. For the $\mu\tau$ channel at heavy Z$^{\prime}$ region, the constraints from the muon collider will be even better than the prospect Belle II results, which shows clear advantages of the future lepton colliders on CLFV searches.

\acknowledgments

This work is supported in part by the National Natural Science Foundation of China under Grant No. 12150005, 12075004, 12175321, 12061141003; the National Key Research and Development Program of China under Grant No. 2018YFA0403900; State Key Laboratory of Nuclear Physics and Technology, Peking University under Grant No. NPT2020KFY04, NPT2020KFY05.


\bibliographystyle{ieeetr}
\bibliography{h}
\end{document}